# The astronomizings of Dr. Anderson and the curious case of his disappearing nova

Jeremy Shears

## Abstract

Dr. Thomas David Anderson (1853 –1932) was a Scottish amateur astronomer famed for his discovery of two bright novae: Nova Aurigae 1891 and Nova Persei 1901. He also discovered more than 50 variable stars as well as making independent discoveries of Nova Aquilae 1918 and comet 17P/Holmes in 1892. At the age of seventy, in 1923, he reported his discovery of a further nova, this time in Cygnus. This was set to be the culmination of a lifetime devoted to scanning the night sky, but unfortunately no one was able to confirm it. This paper discusses Anderson's life leading up to the discovery and considers whether it was real or illusory.

## Introduction

In the early hours of the morning of 9 May 1923, the 70 year old amateur astronomer Dr. Thomas David Anderson (1853 –1932; Figure 1) was carefully examining star fields in Cygnus as part of a nova search programme that he had been conducting for nearly half his life. Suddenly he came upon an unfamiliar star of about the fifth or sixth magnitude. Unfortunately, before he had time to make more detailed observations of the star, his view was obscured by clouds. Nevertheless he sent a telegram to the Royal Observatory Greenwich to report his observation and the approximate location of the object. The following day, photographs taken by the distinguished Greenwich astronomer and a founding member of the BAA, Dr W.H. Steavenson (1894 - 1975), revealed no trace of any new object brighter than about twelfth magnitude. Other photographs taken by astronomers around the world over the course of the next few days also failed to reveal anything unusual. Initially, opinion was divided between those who thought that Anderson had simply made an embarrassing mistake, possibly the misidentification of another field star during a hurried observation by an ageing observer, and those who thought there might be more to Anderson's report, pointing out that he had an international reputation for discovery through his detection of two bright novae, Nova Aurigae 1891 and Nova Persei 1901, and more than 50 variable stars, as well as independently discovering Nova Aquilae 1918 and comet 17P/Holmes in 1892. After all, how could an observer with such intimate knowledge of the night sky have made such an elementary error? After the initial excitement and debate, and in the absence of any further documentary evidence to support the discovery claim, the matter was largely forgotten. But Anderson himself always believed the object was real. Little further was heard from him and he faded into obscurity, his death in 1932 passing unrecorded.

So who was Anderson? And what does modern astrophysics have to say about the events of May 1923? This paper explores these questions.

## Early life

Anderson was an intensely private person with a retiring personality. By contrast to many prominent amateur astronomers of the time who joined national and local astronomical societies, he was not the clubbable type and was neither a member of the BAA nor the RAS. He generally only attended major meetings when invited to receive awards (1). His published works, mainly relating to his nova and variable star discoveries, reveal very little personal information. Most of what is known about his private life came from a series of letters, containing autobiographical notes, which he exchanged with Dr. Hector Copland Macpherson FRSE, FRAS (1888–1956; Figure 2). Macpherson later published the pamphlet *Thomas David Anderson, "Watcher of the Skies*" (2) and a brief note in the BAA Journal (3). Macpherson's work is the source for much of the biographical material I have included in this paper and I suggest that the reader who wishes to gain a deeper insight into Anderson should read Macpherson's original papers. This material is important as it illustrates Anderson's character and establishes the painstaking approach he took in making his discoveries.

Anderson was born on 6 February 1853 in Edinburgh where his father, John Anderson, was a director of an upholstery company. Anderson graduated from Edinburgh University in 1874 with a first-class honours degree in Classics and then went on to train for the ministry at the Scottish Congregationalist College. After completing the course he went on to earn a DSc. in Philology (4) in 1880 with a thesis on "The Latin Conjunctions". He was always destined to be ordained in the Congregationalist church, but on completion of his studies "a grave misfortune" overtook him. "The myopia which I had contracted in my early student days and which had wreaked havoc with the splendid eyesight which I enjoyed when a boy was, I found, increasing to such a degree as to make it absolutely imperative for me to refrain from the writing of sermons. Accordingly, although three congregations offered me the privilege of becoming their pastor, I resolved reluctantly to relinquish the career which I had chosen" (2).

As Macpherson pointed out, the poor state of his eyes, in somebody who later attained fame for his visual discovery of new stars, might have been something of an excuse. Anderson was evidently of a shy disposition and he avoided the limelight as much as possible, generally shrinking from publicity. It may be that his reticence for mixing with people made him feel uncomfortable about embarking on a career in the ministry which by its very essence entails contact with many people. Nevertheless he did undertake temporary preaching duties from time to time (5).

Anderson's first astronomical recollection was at the age of 5 when his father pointed out Donati's comet (6) whilst father and son stood at the front door of their Edinburgh home. At the age of 12 or 13 he purchased a small star atlas and *An Easy Guide to the Constellations,* both by the Scottish author and publisher, Rev. James Gall (1808-1895). With these books in hand he began to learn his way around the night sky.

Having abandoned a career in the ministry, Anderson decided to devote his time to studying astronomy and fortunately he possessed sufficient personal means, inherited from his father, to do so. Years later in a letter to H.P. Hollis (1858-1939) he wrote: (7) "I need hardly say that before the advent of Nova Aurigae my astronomizings were fruitless - fruitless, that is to say, so far as the rest of humanity was concerned, but far from fruitless as regarded myself, for there was for me at least a certain joyful calm when after a long evening spent in writing sermons or in other work I threw up the window and, taking out my little pocket telescope, surveyed the never-failing glory of the midnight sky."

**Nova Aurigae 1891**

Even as a boy, Anderson dreamed of finding a nova. Having read about the appearance of Tycho's star in Cassiopeia in 1572 and Sir John Herschel's suggestion that it might grace the skies once again, he regularly kept an eye on the spot.

Anderson's discovery of Nova Aurigae on 31 January 1892 has become part of astronomical folklore and was re-told in many of the popular astronomy books of the era. The first that anyone heard about the discovery was via an anonymous postcard received the following morning by the Astronomer Royal for Scotland, Ralph Copeland (1837-1905; Figure 3), announcing: "*Nova in Auriga. In Milky Way, about two degrees south of χ Aurigae, preceding 26 Aurigae. Fifth magnitude, slightly brighter than χ*". That evening as it became dark, Copeland turned one of the Observatory's telescopes towards Auriga and confirmed the discovery (8). How Copeland found out that the mysterious discoverer was the modest and unassuming Anderson is not known, but he encouraged Anderson to write a note to *Nature* about the events surrounding the discovery: (9)

"It was visible as a star of the fifth magnitude certainly for two or three days, very probably a week, before Prof. Copeland received my postcard. I am almost certain that at two o'clock on the morning of Sunday, the 24$^{th}$ [January 1892], I saw a fifth magnitude star making a very large obtuse angle with β Tauri and χ Aurigae, and I am positive that I saw it at least twice subsequently during the week. Unfortunately, I mistook it on each occasion for 26 Aurigae, merely remarking to myself that 26 was a much brighter star than I used to think. It was only on the morning of Sunday the 31$^{st}$ [January] that I satisfied myself that it was a strange body….How long before the 24$^{th}$ [January] it was visible to the naked eye I cannot tell, as it was many months since I had looked minutely at that region of the heavens".

Examination of photographic plates taken at various observatories around the world showed that the nova had been visible to the naked eye, yet unnoticed, since 10 Dec 1891 (10) and reached its maximum magnitude of 4.4 on 20 December. A light curve of the nova drawn by BAA member Rev. T.H.E.C. Espin (1858-1934) during the first 6 weeks after its discovery is shown in Figure 4. It was the first nova to be studied by the spectrograph (11).

Only 9 months after discovering Nova Aurigae, Anderson encountered another celestial visitor during one of his nightly vigils. Late in the evening of 8 November 1892, he came across a bright nebulous object in Andromeda (12), which he immediately recognised as a comet. He notified Copeland at the Royal Observatory Edinburgh (13), but soon learnt that the comet had been found two days earlier by Edwin Holmes (1838 or 1839 - 1918) in London. On the evening of 6 November, just before midnight, Holmes decided to take a look at the Andromeda Galaxy, M31, something which he had done regularly since the appearance of a bright new star in 1885 (14). On pointing his 12 inch (30 cm) reflector in the direction of the galaxy Holmes placed his eye to the eyepiece and was shocked by its unusual appearance. Holmes said (15) he "called out involuntarily, *'What is the matter'*? 'There is something strange here.' My wife heard me and thought something had happened to the instrument and came to see." He quickly realised that it wasn't the galaxy, but a bright comet. He notified Maunder at the Royal Observatory, Greenwich, W. H. Maw (1838 – 1924) (16), and Mr. Kidd (Bramley, near Guildford, England). Initially there was some scepticism on Maunder's part that perhaps he had mistaken it for the Andromeda Galaxy. Nevertheless it was confirmed as a new comet, now called 17P/Holmes, on the evening of 7 November. A photograph of the comet taken by E.E. Barnard (1857 – 1923) is shown in Figure 5. It transpired that the comet had passed perihelion nearly five months earlier, but at the time of discovery was undergoing an outburst in apparent brightness, bringing it to naked-eye visibility. It began to fade in the second half of November and a second outburst occurred in mid-January 1893.

**Nova Persei 1901**

At the time of his discovery of Nova Aurigae the only optical aid Anderson had at his disposal was a 1 inch (2.5 cm) pocket telescope. Flushed with success, he resolved to search for other novae and received much encouragement from Copeland to pursue this endeavour. Immediately he purchased "a large binocular" and soon after that a 2¼ inch (6 cm) refractor by Jesse Ramsden. These were supplemented in 1899 with the purchase of a 3 inch (7.5 cm) refractor by William Hume of Edinburgh. With Nova Aurigae under his belt his search for other novae began: (17)

"I worked with might and main, never going to rest as long as the sky remained clear, often rising in the night to see if the clouds had passed away, and, if they had, hurrying downstairs to begin work either with binocular or with telescope. The chief obstacle that I have to contend with in such work is that the only windows in this house from which I can thoroughly examine the heavens face the north-west. Not only is my field of labour thereby greatly circumscribed, my telescope being able to command only that part of the heavens which extends from the equator to +70°, but the discomfort is frequently not inconsiderable, as the northerly and north-westerly winds which so often bring with them transparent, unclouded skies, are in winter and early spring far from being balmy, and can make themselves felt even when the window shutters are partially closed".

Anderson's home at 21 East Claremont Street in Edinburgh's New Town is shown in Figure 6 and is part of a terrace that runs approximately NE-SW, so presumably from his description he did most of his observing from the windows at the rear of the property (Figure 7). Clearly, observing from the open windows must have been less than ideal. Anderson's approach has some parallels to that adopted many years later by George Alcock (1912-2000), the celebrated nova and comet discoverer, who in later life observed the sky with binoculars from various rooms of his home near Peterborough, but in Alcock's case he observed through closed windows thus being spared the chill of the night air.

In the event, Anderson discovered Nova Persei with neither binocular nor telescope. Before retiring to bed on the morning of 22 February 1901 he "was casting a casual glance round the heavens" and at 02.40 UT he found the new star shining at about magnitude 2.7 low in the north-western sky (18). His initial reaction was a feeling of disappointment, for surely someone else must have seen such a bright object beforehand? Nevertheless the following morning he proceeded to the Royal Observatory Edinburgh and on meeting with Copeland shortly after 11 o'clock, learned that he was indeed the first to report the new star (19). Copeland immediately set about despatching telegrams to observatories around the world (20) and not long after dark that evening made his first observation of the nova at 18.30 UT. Of course, being such a bright object there were many independent discoveries. One of the earliest in the UK was at 18:40 UT, by Ivo F.H.C. Gregg (21) of St. Leonards, Sussex, and the well-known variable star observer, and a previous Director of the BAA Variable Star Section, John Ellard Gore (1845-1910) of Dublin, saw it at 23.40 UT (22). Gore and

Anderson were friends and Anderson wrote in a letter to Gore about his feelings on discovering the nova: (23)

"What an absurd sonnet is that in which Keats brackets together the discovery of an ocean and the discovery of a new celestial world. As if the finding of any terrestrial sheet of waters, however large, could be compared for a moment as a source of joy with the first glimpse of a new glory in the already glorious firmament".

Nova Per 1901 (now known as GK Per) was one of the brightest novae of modern times, reaching magnitude 0.2 at its peak.

On 15 July 1901 Anderson was presented with the prestigious Gunning Victoria Jubilee Prize of the Royal Society of Edinburgh for his discovery. Rev. Prof. Robert Flint (24) (1838 – 1910), who chaired the meeting, informed the assembled fellows: (25)

"the value of Dr. Anderson's timely discovery is enhanced by the fact that it afforded astronomers the unique opportunity for watching the course of development in the *initial* stages of this phenomenon, and in this respect the importance of the discovery has been fully appreciated by astro-physicists"

The following year he was awarded the Jackson-Gwilt Medal of the Royal Astronomical Society. He travelled down to London to receive the medal at the February meeting from the RAS President J.W.L. Glaisher (1848-1928; Figure 8) (26), who noted: (27)

"It is no small matter to have discovered one of these Novae, but it is a veritable *tour de force*, such as *à priori* would have seemed almost incredible, to have discovered both, and I am delighted….to congratulate you on your success and do honour to your astronomical zeal and intimate knowledge of the sky"

At the same meeting the Dutch astronomer J.C. Kapteyn (1851 – 1922; Figure 9) (28) was awarded the RAS Gold Medal. Whilst Kapteyn and his wife made the most of their visit to England by staying for a few days and visiting the sights, Anderson, for whom the journey to London was already an ordeal, not least because of his impaired hearing, "returned to his 'star-gazing' in Edinburgh very promptly. The big telescopes at Greenwich and elsewhere, he said, did not appeal to him anyway" (29). Prof. H.H. Turner (1861-1930) noted Anderson's comments to him after the presentation of the award: (29)

" 'I'm not an astronomer', he said; I am an astro*phil*; whenever the stars are shining I *must* be looking'…..he knows the stars so well, especially near the Milky Way, that he does not think one could appear brighter than the sixth or perhaps the seventh magnitude without his detecting it"

**Variable Star Discoveries**

During Anderson's search for novae he also encountered previously unknown variable stars, of which he discovered at least 52 in total between 1893 and 1908 and which are listed in Table 1. Even without any nova discoveries to his credit, he would still have been well-known as a discoverer of variable stars. His first discovery was the long-period variable V Cas on the night of 8 December 1893 which he announced in *Astronomische Nachrichten* (30). In a letter to Macpherson many years later (2), he wrote "It was with a binocular that I saw V Cassiopeiae shining as a star of the eighth magnitude in a place where I had never before noticed a star, and where the Bonn chart showed only one of the 9½ magnitude. Subsequent examination proved that it was a variable star, then near maximum. The others down to 1898 were discovered with my 2¼ inch refractor, since then with the 3 inch one". A light curve of V Cas based on observations from the BAA Variable Star Section database is shown in Figure 10.

The month before his discovery of V Cas, he had also reported that another star listed in the Bonner Durchmusterung as being of magnitude 8.7, but was apparently "missing". Correspondence with Sir Robert Ball (31) (1840-1914) and Karl Friedrich Küstner (32) (1856-1936) revealed further observations of the star at various different magnitudes, thus showing it to be variable ion brightness (33). The star is another long-period variable now known as T And.

Anderson often reported his discoveries to Copeland, with whom he developed a spirit of close co-operation, who arranged follow-up observations with the telescopes at the Royal Observatory on Calton Hill (34). Anderson's last reported variable star discovery was RW Aqr, reported on 31 October 1908. Quite why he stopped reporting discoveries at that point is not known, as he continued to observe the skies in the search for novae and must surely have come across variables, but he apparently viewed his variable star research and nova search as independent activities (35). It might have been that after 52 variable star discoveries, and with large numbers of discoveries now being made by the photographic sky surveys, he simply lost interest in reporting them. Copeland had died 3 years previously, on 27 October 1905, and whether this was a contributory factor as a result of Copeland no longer being able to confirm Anderson's discoveries, or to provide encouragement, must remain speculation. Forty-seven discoveries were made before Copeland's death and only 5 subsequently.

**Nova Aquilae 1918**

In early 1905, Anderson moved from Edinburgh to Northrig, near Haddington, a rural area some 20 miles (32 km) east of Edinburgh (36). The main reason appears to have been the installation of electric lights in East Claremont Street which interfered with his observing. In 1910 he moved again, this time to Thurston Mains, Innerwick, a further 14 miles (22 km) to the east.

Anderson's next appearance on the public stage, after the paper announcing the discovery of RW Aql in 1908, was some ten years later in connexion with the appearance of the brightest nova of the 20th century in June 1918, Nova Aquilae 1918. Anderson's independent discovery of the nova was reported in the *Scotsman* newspaper: (37)

"The new star in Aquila has not escaped the observation of Dr. Thomas D. Anderson, who is known to astronomers throughout the world as the discoverer of two famous 'new stars'. Yesterday we received the following telegram from Dr Anderson:-

> '*Thurston Mains, Innerwick, East Lothian.*
> *A new star of the first magnitude, as bright as Vega, has burst out in the constellation of the Eagle.*
> *Its right ascension is 18 hours 45 minutes; and its declination is half a degree north.*
> *It shines at present with a blue light*' "

There were numerous independent discoveries of the nova, which is now known as V603 Aql, as darkness fell around the world on 8 June 1918. The first person to have seen it from Britain was probably Miss Grace Cook, a BAA member observing from Stowmarket, at 21.30 UT (38), although the first reliable visual detection was by G. N. Bower observing from Madras (39) in India, some 5 hours earlier at 16.30 UT (40).

**A nova in Cygnus?**

After Anderson's brief mention in print at the time of the appearance of Nova Aquilae 1918, nothing further was heard from him until early May 1923, when a telegram was received at the Royal Observatory Edinburgh on Blackford Hill by Prof. R.A. Sampson (1866 – 1939; Figure 11), who was Astronomer Royal for Scotland at the time: (41)

*"Nova Cygni, half a degree north, following 70 Cygni, fifth magnitude. Brighter than 70, but fainter than 72. Rough estimate of position with a binocular. Right ascension 21 degs., 25 m. 25 s. North declination 37 degs., 6 m. Anderson, Thurston Mains, Innerwick"*

Anderson described the events of the night in more detail in a later letter to Sir Frank Dyson (1868 – 1939), who was Astronomer Royal (42). Anderson had been scanning the skies before retiring to bed when at about 00.40 UT on 9 May 1923, when to "my astonishment…on coming to the place of 69 and 70 Cygni, I saw instead of two stars, three arranged in a straight line, except that B (the middle one) was slightly to the south-following side of such a line; and the distance of each star from its nearest neighbour was almost exactly half a degree, but the distance of A (the most northerly) from B was slightly less than that of B to C. They were stars of the fifth to the sixth magnitude, A being brighter than B, and B than C".

At this point it appears that Anderson made an elementary procedural error that was to cost him dear, as he was soon to find out:

"After receiving the shock inevitable on occasions when there is a great upset in nature, I did a stupid thing. I hurried at once into the house from the pathway outside where I had been standing, in order to note the exact time of my observation…instead of ascertaining at once by means of the adjacent fainter stars which of the three were 69 and 70 and which was the newcomer. When I went outside again to scan the heavens, all of the following half of the Swan was, alas ! covered with clouds, which drifted along from the south in a long weary cavalcade"

Having considered the matter further whilst consulting a number of star atlases and catalogues he "came at last to the conclusion that A was the nova" and his telegram to Sampson, as well as a further one to the Royal Observatory, Greenwich, was based on this position. However, upon further reflection he changed his mind and concluded that he was "tolerably certain that C was the nova" and sent further telegrams with the corrected position, which, we precessed to J2000.0 co-ordinates is RA 21h 26 min 16 sec Dec. 36° 24'.

Having received Anderson's telegram, W.H. Steavenson at Greenwich used a 3 inch (7.5 cm) portrait lens with a focal length of 13 inches (33 cm) to secure a photograph the region around 69 and 70 Cyg on the following morning, 10 May. The 3° field contained both of the positions notified by Anderson, the original one and the corrected one, but there was no sign of any new object brighter than 12$^{th}$ magnitude (43). Other observatories around the world also failed to confirm the nova. Visual and photographic searches at Yerkes Observatory on the night of 12 May yielded negative results (44). Plates from Harvard College Observatory on 13 May showed nothing unusual brighter than $11^{th}$ magnitude (45) and photographs from the Uccle Observatory in Brussels on 14 May showed nothing brighter than magnitude 9.5 (46).

In the absence of confirmatory observations, discussions soon began about possible explanations during the RAS meeting held in London two days later, 11 May (47). Steavenson was of the opinion that Anderson had simply mistaken 69 or 70 Cyg for the nova, an idea supported by Rev. T.E.R. Phillips (48) (1868-1942). Phillips also pointed out that his own copy of Proctor's star atlas omitted 69 Cyg and speculated that Anderson might also have an atlas with the same star missing and that the nova was actually 69 Cyg. Of course, the fact that Anderson specifically mentioned 69 Cyg shows that he was well aware of its existence. Moreover, Anderson was certain that he saw three stars: both 69 and 70 Cyg as well as the nova. The RAS President J.L.E. Dreyer (1852 – 1926; Figure 12), who was personally acquainted with Anderson, also thought that a misidentification of a normal field star was the most likely explanation. However, Prof. R.A. Sampson, who also knew Anderson and had received his discovery announcement, could not quite accept this view, pointing out to the meeting audience that Anderson was the discoverer of two novae and was the recipient of the RAS's Jackson-Gwilt medal. Sampson went on to say "he is a keen observer, and is said to know the entire visual heavens by heart. That he made a mistake seems to me to be most improbable, and I fail to understand the matter". However, although a mistake was "most improbable", Sampson did allow for it being possible: "If a mistake has been made, we must recall that he is now an old man, and perhaps relied upon his memory beyond what was justified".

As the days passed, a consensus of opinion developed that Anderson had indeed made a mistake. H.P. Hollis writing in the English Mechanic (49), whilst acknowledging Anderson's previous discoveries, concluded that the events "scarcely leave room for any doubt that a mistake was made somewhere, and that the nova had not appeared". Only H.H. Turner, Savilian Professor of Astronomy in the University of Oxford and a great admirer of Anderson, continued to give Anderson the benefit of the doubt: (50)

"With a less experienced observer [than Anderson] we might have suspected some mistake, but Anderson's knowledge of the heavens is too comprehensive and minute to allow of this interpretation. We may accept from him without hesitation the statement that he saw a nova".

Anderson himself continued to believe that he has seen a nova, although he conceded that it must have "faded away with miraculous rapidity" (42).

**An embarrassing mistake, or something more?**

Although conditions were apparently good at the moment Anderson came upon the object in Cygnus, they rapidly deteriorated with advancing cloud that prevented him making more detailed confirmatory observations. There is no doubting Anderson's intimate knowledge of the night sky. Copeland believed that Anderson knew the sky so well that he could detect a new fifth magnitude star in almost any part of the heavens (51). And of course his track record speaks for itself through his discovery of two novae, some 52 variable stars, and independently discovering a third nova plus a comet. But nobody is infallible and mistakes can be made by even the most experienced amateur and professional astronomer from time to time, so the likelihood of error must be accepted. It is also true that the eye is easily misled and an observation subject to misinterpretation via the imagination. Could Anderson simply have been confused whilst observing this star rich region of Cygnus? Did advancing age and fading memory contribute, as suggested by Sampson? The generally accepted view was that Anderson had simply made a mistake on the morning of 9 May 1923, misidentifying a known star for a nova.

At the same time other explanations for Anderson's observation cannot be excluded. As the historian of astronomy Richard Baum has pointed out (52), we must be very careful about casually dismissing apparently anomalous observations, especially those made by reputable observers, simply because they are difficult to explain or they don't fit the conventional wisdom of the time. Indeed the recent observations of impacts on Jupiter (53) have taught us to be more circumspect. New information or understanding surfacing in a later epoch can shed new light on earlier observations. For example, BAA Comet Section Director, Jonathan Shanklin, has mentioned the possibility that bright objects that have been seen near the sun by many observers over the years, but especially in the nineteenth century, might actually be Kreutz "sun-grazing" comets in spite of the scepticism with which these reports were received at the time (54). Shanklin also points out that some of the Kreutz comets show no tail at all and it is possible that some supposed observations of Vulcan were actually tiny Kreutz group comets. It is only in recent years, thanks to the SOHO and STEREO missions (55), that the astronomical community has become aware of how common bright sun-grazing comets actually are, as most pass the dazzlingly bright limb of the sun unnoticed by Earth-bound observers.

So let us for a few moments take Anderson's report at face value and consider that he had indeed seen a star-like object of the fifth or sixth magnitude, which faded to below $12^{th}$ magnitude within 24 hours, an "optical transient" in modern parlance. What could it have been? Anderson himself believed that he had seen a very-fast declining nova, faster than any seen before, an explanation which was also entertained by Prof. H.H. Turner (50). However, a decline of some 6 magnitudes in brightness in a day is unknown amongst classical novae, where decay times are normally of the order of days to weeks, so this idea can safely be rejected.

Maybe the explanation for Anderson's optical transient lies with one of the most exotic classes of objects known to modern astrophysics: the *Gamma Ray Bursters* (GRBs). Optical transients associated with GRBs were first observed in the late 1990s (56). Optically, the brightest GRB yet observed was GRB 080319B in 2008, which peaked at magnitude V = 5.3 and might have been visible with the naked eye for half a minute to somebody looking in the right direction (57). Decay times of GRBs are of the orders of hours to a few days (58). Richard Strom, of the Netherlands Institute for Radio Astronomy, and his colleagues have postulated (59) that significant numbers of naked eye GRBs could have been observed in the course of human history and suggest that historical records should be reviewed to see whether any can be identified.

Apart from GRBs other astronomical phenomena can create star-like transients that could conceivably become bright enough for naked-eye detection (60). *Superflares* on ordinary solar-type stars cause a brightness increase of the order of a magnitude over a few minutes (61), with a decay on a similar timescale, but such an amplitude is modest compared to Anderson's nova. *Gravitational microlensing events* can cause stars to brighten by several magnitudes over a period of a few days. For example, in October 2006 an unremarkable star in Cassiopeia brightened from V = 11.4 to V= 7.5 in about a week, with a symmetrical fade (62). Again the amplitude and speed of decline are relatively modest compared with Anderson's nova, but it is possible that higher amplitude and faster events might occur.

Anderson's observation is certainly not the only transient object to have been discussed in the historical record and we shall now turn to another in which Anderson himself took a particular interest.

**Hertzsprung's Enigmatic Object** (63)

In 1927 the Danish Astronomer Ejnar Hertzsprung (1873-1967; Figure 13), of Hertzsprung-Russell diagram fame, was examining plates in the archives at Harvard College Observatory. Whilst carrying out photometry in the field of RX Cas he found a fairly bright image on a plate which had been exposed on 15 December 1900, but which was absent from other plates. He noted (64) that "the reality of the object seems practically settled by the fact that there are two exposures of similar duration on the plate and the image in question is double in the same way as those of the stars". The two images, taken within an hour of each other, were round and slightly nebulous and in the second image the object was about 0.7 magnitudes brighter. Careful study of the plate allowed Hertzsprung to rule out any defects in the emulsion and no trace of a similar object was found on plates taken either the night before or the night after 15 December; several hundred other photographs in the archive also showed nothing unusual. He concluded that the images were of an unknown celestial phenomenon.

In 1951, Dorrit Hoffleit (1907-2007) suggested (65) that the object might have been a flare star, although she pointed out that the flare would have been of abnormally large amplitude. Flare stars were unknown at the time of Hertzsprung's discovery (66). Hoffleit searched through about 3000 Harvard photographs of the region that had been obtained since Hertzsprung's investigation, but she too found nothing unusual.

It is interesting that the last communication from Anderson that appeared in print was a letter he wrote to *The Observatory* in August 1927 (67) on the subject of Hertzsprung's enigmatic object. Recalling his observations of the outburst of Comet 17P/Holmes, which, as we noted earlier, Anderson had independently discovered in November 1892, he wondered whether the object that Hertzsprung had found was also a comet in outburst. Hertzsprung wrote in reply: "I think your suggestion is the best one which has been made so far".

However, the story did not end there. Bradley Schaefer examined Hertzsprung's original plate and found several other nebulous objects upon it (68). He concluded that they were plate defects and found them to be similar in appearance to defects seen on other Harvard plates. Thus Hertzsprung's enigmatic object was not a celestial phenomenon after all.

Nevertheless it is interesting to contrast the events surrounding Anderson's report of Nova Cyg 1923 and Hertzsprung's enigmatic object. Both were short-lived isolated appearances of a bright object that defied obvious explanation. But there were two key differences. Firstly, there was the question of the credibility of the discoverers. Hertzsprung was an eminent professional astronomer at the height of his career; by contrast Anderson was by then an aged and largely forgotten amateur astronomer. Secondly, there was the credibility of the evidence itself. Hertzsprung apparently had documentary evidence in the form of a photograph showing his object, even though we now know that it was a defect. This could be pored over and analysed by anyone who was interested. By contrast poor Anderson "only" had his uncorroborated visual observation, which by his own admission he had made hurriedly. Moreover his ambiguity of recollection of which was the nova and which were field stars and his subsequent change in the position he gave for the object, suggested confusion which probably didn't help his case.

**Epilogue**

During 1926 Anderson moved from Thurston Mains to Stuartslaw farmhouse, at Edrom in the Scottish Borders where he continued observing (69) (Figure 20). Two years later, he suffered a stroke from which he made a complete recovery. During his convalescence he studied Danish and Russian, becoming sufficiently competent to read books in both languages. Once back to full health, he could be found tending his vegetable garden by day and "watching the heavens at night". He passed away on 31 March 1932, aged 79 years.

Anderson will be long remembered for his discoveries of Nova Aurigae and Nova Persei. These objects were important because they were amenable to study by the rapidly developing technique of spectroscopy, one of the most powerful tools of the emerging science of astrophysics, which allowed insight to be gained into the underlying astrophysics of nova explosions. But the discoveries are equally interesting from a human perspective and are compelling in their own right. Here we have a reclusive man, whose passion was to study the skies each clear night with minimal optical aid and whose dedication and persistence were rewarded by the discovery of two bright novae, and yet whose modesty prevented him seeking personal glorification. This surely is testimony to the strength of the human spirit! It is therefore not surprising that these stories of discovery were told and retold in countless publications at the time and right up to the present day.

Anderson's discoveries are also relevant to amateur astronomers of today. Whilst professionals have access to large telescopes, the time available to them is limited and strictly controlled by committees that allocate time depending on the merit of the research. By contrast, the amateur, albeit with more modest means at his disposal, but with time and dedication on his side, can still make a real contribution to astronomy. Even in an age of large sky surveys, amateur astronomers continue to discover novae, comets and supernovae.

We shall never know Anderson's personal feelings about the response to his "disappearing" nova. Clearly he remained convinced that what he saw on the morning of 9 May 1923 was real, but did he simply shrug off the incredulity with which his report was received by the majority of astronomical community? Even those who were more supportive of the idea that he had made a real discovery were happy to let the story drop, perhaps through fear of damaging his reputation. However, even if the observation had been illusory, the episode in no ways tarnishes the memory of this remarkable "watcher of the skies".

**Acknowledgements**

I am indebted to Richard Baum for his enormous encouragement and for providing insight into apparently anomalous observations from the astronomical record book. Professor Bradley Schaefer (Department of Physics and Astronomy, Louisiana State University) brought his work on the Hertzsprung's enigmatic object to my attention, in which he showed that it was not a celestial phenomenon, but a plate defect. Vicki Hammond, Journals and Archive Officer at the Royal Society of Edinburgh, kindly researched references to Anderson in the Proceedings of the RSE and made copies available to me. I thank Richard London and Kate Snowden of Simpson & Marwick, Edinburgh (www.edinburghprimeproperty.com) for providing the photographs of 21 East Claremont St., Edinburgh.

This research made use of data from the BAA Variable Star Section database contributed by observers worldwide, the NASA/Smithsonian Astrophysics Data System, the AAVSO Variable Star Index and SIMBAD, operated through the Centre de Données Astronomiques (Strasbourg, France). I also made extensive use of two wonderful resources for the astronomical historian: scanned back numbers of the BAA *Journal*, which are available largely thanks to the efforts of Sheridan Williams, and the *English Mechanic*, courtesy of Eric Hutton.

**Address**

"Pemberton", School Lane, Bunbury, Tarporley, Cheshire, CW6 9NR, UK
[bunburyobservatory@hotmail.com]

**References and notes**

1. Although Anderson rarely attended meetings, he was present at the second meeting of the newly formed Astronomy Institution of Edinburgh on 17 February 1904, according to the meeting report in the English Mechanic, 2031, 54 (1904).

2. Macpherson H.C., Publications of the Astronomical Society of Edinburgh - No. 2. Printed by T. and A. Constable, printers to the University of Edinburgh (1955). The pamphlet contains the substance of Macpherson's Presidential address to the. Astronomical Society of Edinburgh on 6 November 1953. Macpherson also published the definitive "Biographical Dictionary of Astronomers" in 1940, but this did not have an entry for Anderson, an omission which Macpherson wanted to put right via the pamphlet.

3. Macpherson H.C., JBAA, 50, 302-303 (1940). Macpherson had also written a brief note on "Two Scottish Astronomers of Today", Anderson and Alexander William Roberts (1857-1958), in PA, 16, 397-403 (1908).

4. Philology is the study of language in written historical sources and as such it combines literary studies, history and linguistics.

5. Macpherson noted that although frequently referred to as "the Rev. Dr. Anderson" or a "Scottish clergyman", this was not strictly the case. In the Congregationalist church the title of Reverend is not endowed until the person is ordained to a charge and. Anderson was never ordained.

6. C/1858 L1 was one of the brightest comets of the nineteenth century.

7. Hollis H.P., Observatory, 25, 124-127 (1902). Hollis quotes from a letter written to him by Anderson following the discovery of Nova Persei 1901.

8. Copeland noted that "Already on the night of February 1, a small spectroscope revealed the presence of bright lines, of which some account was at once sent to the Central Station for Astronomical Telegrams at Kiel, and also to the President of the British Association, Dr. W. Huggins. Already in the daytime, before the star was visible, a message had been forwarded to Greenwich Observatory", Copeland R.A., Trans. Roy. Soc. Edinburgh, 37, 51-58 (1892).

9. Anderson T.D., Nature, 45, 365 (1892). Anderson's letter was written from his home at 21 East Claremont Street, Edinburgh, on 13 February 1892 and published in Nature on 18 February 1892.

10. The nova was found to be present on a photograph by E.C. Pickering (1846-1919) taken from Harvard College observatory on 20 December 1891. By contrast, a photograph by Max Wolf (1863-1932) at the Heidelberg Observatory showed nothing brighter than eighth magnitude.

11. The Recurrent Nova T CrB had been observed visually with the spectroscope in 1866, but Nova Aurigae 1891 was the first to have its spectrum photographed.

12. Copeland R., AN, 131, 133 (1892).

13. Not long after the discovery of Nova Aurigae, Copeland had narrowly escaped from a fire at their residence at 15 Royal Terrace, Edinburgh, on 8 March 1892. "The building, which is Government Property, was destroyed, and Dr. Copeland, with his family, only escaped by means of an extemporized rope of sheets and blankets, whereby they descended from their bedrooms to a balcony in front of the house", JBAA, 2, 201 (1892).

14. The new star, S And, was discovered on 19 August 1885 by Isaac Ward (1834-1916) of Belfast. It is now known to have been a supernova, the first ever discovered outside the Milky Way.

15. Holmes E., Observatory, 15, 441-443 (1892).

16. W.H. Maw was the BAA's first treasurer, serving between 1890 and 1912.

17. Anderson T.D., Observatory, 25, 124-127 (1902).

18. Further details of the circumstances surrounding the discovery can be found in H.H. Turner's "Astronomical Discovery" (publ. Edward Arnold, London, 1904). Ralph Copeland's account is in English Mechanic, 1875, 54 (1901).

19. Prof. H.H. Turner, who knew Anderson, gives a slightly more colourful account of Anderson's visit to Copeland to announce the discovery in his book "Astronomical Discovery". "More in a spirit of complaint than of inquiry, he made his way to the Royal Observatory of Edinburgh next day to hear what they had to say about it, though he found it difficult to approach the subject. He first talked about the weather, and the crops, and similar topics of general interest; and only after some time dared he venture a casual reference to the "new portent in the heavens". Seeing his interlocutor look somewhat blank, he ventured a little farther, and made reference to the new star in Perseus; and then found to his astonishment, as, also to his great delight, that he was the first to bring news of it".

20. Copeland R., Proc. Roy. Soc. Edinburgh, 23, 365-369 (1901).

21. Gregg was not sure quite what to make of the object so he sent a telegram to E.E. Markwick (1853 – 1925), then Director of the BAA Variable Star Section. Markwick had just finished dinner at the time, and was enjoying a postprandial cigar with a fellow Army Officer, but the two immediately went outside to see the new star. Markwick continued to observe it throughout the evening.

22. Gore had spent the evening with a friend in Dublin and was returning home when he looked up and saw the nova. His observations of the nova are described in a biography of Gore which I have prepared for this Journal: Shears J., JBAA, Accepted for publication (2012).

23. Gore quotes from this letter in: Gore J.E., Studies in Astronomy, publ. Chatto & Windus, London, 1904.

24. Robert Flint was a Scottish theologian and philosopher. From 1876 to 1903 he was Professor of Divinity at the University of Edinburgh. Unfortunately Copeland was unable to be present due to illness.

25. Proc. Roy. Soc. Edinburgh, 23, 448-450 (1901).

26. James Whitbread Lee Glaisher, FRS FRAS, spent his career lecturing at Cambridge University. He was also a prolific author, writing articles on astronomy, special functions, numerical tables and number theory. He served as President of the London Mathematical Society from 1877-86, and of the Royal Astronomical Society twice, in 1886-88 and 1901-03.

27. MNRAS, 62, 342-343 (1902).

28. Jacobus Cornelius Kapteyn was Professor of Astronomy at the University of Groningen.

29. Quoted from H.H. Turner's "From an Oxford Notebook" column in the Observatory March 1902: Observatory, 25, 141-142 (1902).

30. Anderson T.D., AN, 134, 211 (1894).

31. Sir Robert Stawell Ball was appointed Lowndean Professor of Astronomy and Geometry at Cambridge University in 1892, at the same time becoming director of the Cambridge Observatory. He wrote many popular articles and books on astronomy.

32. Küstner was a German astronomer who also made contributions to Geodesy. In 1888 he reported the Polar Motion of the Earth.

33. Anderson T.D., Nature, 49, 101 (30 November 1893).

34. An example was U And, which Anderson had been following as a newly discovered variable in early 1895. When it faded below the detection limit of Anderson's equipment, Copeland arranged follow-up observation with the 12 inch (30 cm) and 24 inch (60 cm) reflectors on Calton Hill. See: Copeland R., AN, 139, 116-117 (1895). Anderson announced the discovery in the same edition of AN, as well as his discovery of W Peg: Anderson T.D., AN, 139, 117 (1895).

35. H.H. Turner, whom as has been mentioned earlier knew Anderson, noted that Anderson's "discoveries [of novae] were the outcome of a deliberate and independent research, not connected with his work on variables". See: Turner H.H., Observatory, 25, 141-142. (1902). Anderson was also interested in stars which might have changed in brightness since medieval or classical times. For example, in an article in "Knowledge" magazine in July 1893 he considered whether θ Eri might have faded since the time of Ptolemy.

36. Macpherson's biography of Anderson's says the move was in the Spring of 1904. However, Anderson was still writing from his East Claremont address, announcing variable star discoveries, as late as 23 Feb 1905. The next announcement was from Northrig on 5 May 1905.

37. The Scotsman, Tuesday 11 June 1918, pg 4. The article suggests that Anderson's telegram was received by the paper on 10 June, which might mean he did not see the nova until the night of the 9 June. However, on the night of the 9 June various observers reported that the nova was much brighter that Vega, which Anderson surely would have remarked upon, suggesting he had actually seen the star on the evening of 8 June. Macpherson, in his biography of Anderson, also notes he found the star on the, 8 June.

38. Cook A.G., JBAA, 28, 208 (1918).

39. Madras is now known as Chennai.

40. Evershed J., Nature, Volume 102, Issue 2554, pp. 105-106 (1918). Observers located at more easterly longitudes were favoured. The nova was apparently detected at about the same time as Bower by Radha Gobinda Chandra (1878 - 1975), a BAA member observing from Bagchar, India (see Biswas S.N, Mukhopadhyay U. & Ray S., A Village Astronomer: Life and Works of R. G. Chandra at http://arxiv.org/pdf/1102.2383.pdf).

41. The contents of the telegram are quoted in "Scientific News" in English Mechanic, 3034, 18 May 1923. Sampson was Astronomer Royal for Scotland 1910–1937; he had succeeded Frank Dyson (served 1905–1910), who in turn had succeeded Copeland when he died in 1905.

42. Anderson T.D., Observatory, 46, 198-199 (1923). Anderson wrote the letter on 15 May 1923, some four days after the discovery.

43. Steavenson presented his photograph at the RAS meeting on Friday 12 May 1923: Observatory, 46, 173-175 (1923).

44. "Reported Nova in Cygnus": Pop. Astron., 31, 421-422 (1923).

45. Harvard College Observatory Bulletin 786, 14 May 1926.

46. Staff at Uccle received notification of the nova on 12 May, but cloud prevented observations until 14 May: Delporte E., Ciel & Terre, 39, 127 (1923).

47. Observatory, 46, 173-175 (1923).

48. Rev. Theodore Evelyn Reece Phillips was a renowned planetary observer. He was Director of the BAA Jupiter section from 1900–1933 and Director of the Saturn section from 1935–1940. He was President of the BAA 1914-1916 and of the RAS 1927–1929.

49. Hollis H.P., English Mechanic, 3035, item 169 (1923). "Letters to the Editor" in the edition of 25 May 1923.

50. Turner H.H., Observatory, 46, 233 (1923). Turner made these comments in the July 1923 edition of "The Observatory" in his long-running series of notes on contemporary astronomy, "From an Oxford Notebook".

51. English Mechanic, 1876, 74 (1901).

52. Baum R., The Haunted Observatory: Curiosities from the Astronomer's Cabinet, publ. Prometheus Books (2007). Richard Baum's beautifully written book considers many apparently anomalous observations from the historical record.

53. Some 15 years after the bombardment of Jupiter by comet Shoemaker-Levy 9, which caused visible impact scars, a new impact scar appeared on 2009 July 19, discovered by Anthony Wesley from New South Wales, Australia. See: Rogers J., JBAA, 119, 235 (2009).

54. Shanklin J. Personal communication (2012).

55. SOHO is the Solar and Heliospheric Observatory and STEREO is the Solar Terrestrial Relations Observatory.

56. van Paradijs J., Groot P.J., Galama et al., Nature, 1997, 386, 686 (1997).

57. Racusin J.L. et al., Nature, 455, 183 (2008).

58. Rau A., PASP, 121, 1334 (2009). This paper discusses strategies for discovering optical transients, including GRBs, via the "Palomar Transient Factory" project.

59. Strom R.G., Zhao F. and Zhang C., Research in Astronomy & Astrophysics, 12, 260-268 (2012) available at arxiv.org/abs/1110.5970.

60. Lior Shamir and Robert Nemiroff considered the gamut of known, extra-terrestrial, astronomical phenomena that could create transients that become bright enough for naked-eye detection. These include the phenomena discussed in the present paper (novae, GRBs, stellar superflares, microlenses) as well as supernovae (again, Anderson's object faded too rapidly to be a conventional supernova). Shamir L. & Nemiroff R.J., AJ, 138, 956 (2009).

61. Schaefer B.E., King J.R. and Deliyannis C.P., AJ, 529, 1026-1030 (2000).

62. Gaudi B.S. et al., AJ, 677, 1268-1277 (2008). The results presented in the paper include a large number of data obtained by amateur astronomers in the informal global network of observers equipped with small telescopes and CCDs known as the Center for Backyard Astrophysics.

63. "Hertzsprung's Enigmatic Object" was the title of one of Joseph Ashbrook's (1919 - 1980) Astronomical Scrapbook columns in Sky & Telescope in which he described Hertzsprung's discovery in detail. See: Sky & Telescope, 34, 382-383 (1962).

64. Hertzsprung E., Bulletin Harvard College Observatory, 845, 3-5 (1927).

65. Hoffleit D., Bulletin Harvard College Observatory, 920, 32 (1951).

66. The earliest confirmed observations of flare stars are attributed to W.J. Luyten (1899-1994), who discovered strongly variable spectra and brightness in two high proper motion stars, V1396 Cyg and AT Mic. The archetypal flare star, UV Cet, was discovered in September 1948.

67. Anderson T.D., Observatory, 50, 249 (1927).

68. Schaefer B.E., PASP, 95, 1019-1020 (1983).

69. In Anderson's time the county was Berwickshire.

70. Espin T.H.E.C., JBAA, 3, 328 - 332 (1892).

71. Ball R.S., A Popular Guide to the Heavens (1905).

| Star | | Year of announcement | Type | Range (mag.) | Period (d) |
|---|---|---|---|---|---|
| RR | And | 1901 | M | 8.4-15.6 | 330 |
| T | And | 1893 | M | 7.7-14.5 | 281 |
| U | And | 1895 | M | 9.0-15.0 | 347 |
| V | And | 1896 | M | 9.0-15.2 | 256 |
| W | And | 1899 | M | 6.7-14.6 | 397 |
| X | And | 1900 | M | 8.5-12.5 | 343 |
| Y | And | 1900 | M | 8.2-15.1 | 220 |
| RT | Aql | 1897 | M | 9.6-14.5 | 327 |
| RU | Aql | 1898 | M | 8.7-14.8 | 274 |
| RV | Aql | 1900 | M | 8.1-15 | 218 |
| RV | Aqr | 1907 | M | 9.0-<13 | 453 |
| RW | Aqr | 1908 | M | 8.5-14.5 | 140 |
| X | Aur | 1900 | M | 8-13.6 | 163 |
| Z | Aur | 1903 | SRD | 9.2-11.7 | - |
| RT | Boo | 1907 | M | 8.3-13.9 | 274 |
| RR | Cas | 1900 | M | 8.5-14.7 | 300 |
| V | Cas | 1893 | M | 6.9-13.4 | 229 |
| Z | Cas | 1898 | M | 8.5-15.4 | 496 |
| W | CrB | 1902 | M | 7.8-14.3 | 238 |
| T | CVn | 1897 | SRA | 8.9-11.7 | 290 |
| SX | Cyg | 1899 | M | 8.2-15.2 | 411 |
| TZ | Cyg | 1901 | LB | 9.6-11.7 | - |
| X | Del | 1895 | M | 8.0-14.8 | 282 |
| Y | Del | 1902 | M | 8.8-16.5 | 468 |
| U | Dra | 1897 | M | 9.1-14.6 | 316 |
| V | Dra | 1900 | M | 9.5-14.7 | 278 |
| R | Equ | 1900 | M | 8.7-15 | 261 |
| X | Gem | 1897 | M | 7.5-13.8 | 264 |
| RS | Her | 1895 | M | 7 - 13 | 219 |
| RT | Her | 1896 | M | 8.5-15.5 | 298 |
| RU | Her | 1896 | M | 6.7-14.3 | 484 |
| RV | Her | 1897 | M | 9-15.5 | 205 |
| RY | Her | 1899 | M | 8.3-14.1 | 221 |
| SS | Her | 1901 | M | 8.5-13.5 | 107 |
| S | LMi | 1904 | M | 7.5-14.3 | 234 |
| S | Lyn | 1897 | M | 8.5-14.8 | 296 |
| T | Lyn | 1906 | M | 8.8-13.5 | 406 |
| V | Lyr | 1895 | M | 8.2-15.7 | 374 |
| W | Lyr | 1896 | M | 7.3 - 13.4 | 198 |
| RS | Mon | 1904 | M | 9.1-15.0 | 263 |
| RT | Mon | 1905 | SRB | 8.1-10.3 | 107 |
| RT | Oph | 1901 | M | 8.6-15.5 | 426 |
| RX | Oph | 1905 | M | 9.8-<13 | 323 |
| RY | Oph | 1905 | M | 7.4-13.8 | 150 |
| SS | Oph | 1907 | M | 7.8-14.5 | 181 |
| RR | Peg | 1901 | M | 8.5-14.9 | 264 |
| RT | Peg | 1902 | M | 9.4-15.4 | 215 |
| W | Peg | 1895 | M | 7.6-13 | 345 |
| X | Peg | 1898 | M | 8.8-14.4 | 201 |
| Y | Peg | 1900 | M | 8.9-16.4 | 207 |
| Z | Tau | 1900 | M | 9.8-18.0 | 466 |
| V | UMa | 1901 | SRB | 9.5-11.5 | 207 |

Table 1: Anderson's variable star discoveries

Data on type, magnitude range and period are from the AAVSO Variable Star Index.

M= Mira type long period variable; SRA, SRB, SRD = semi-regular variables, LB = irregular variable

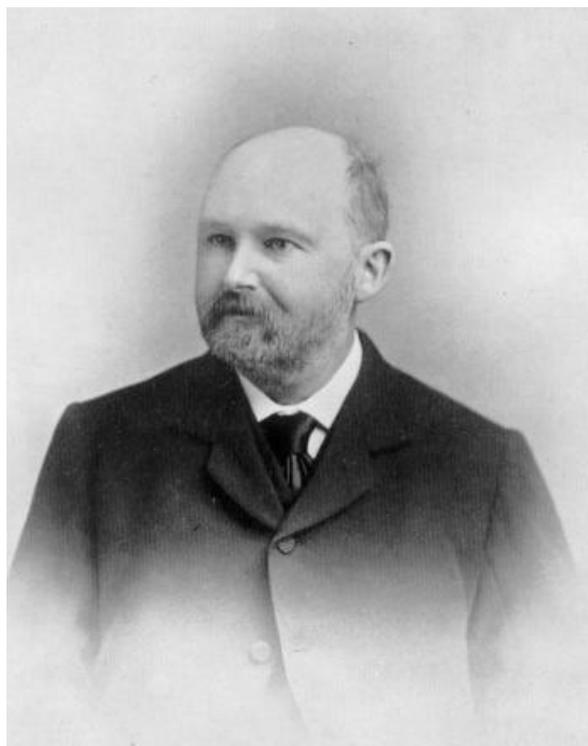

Figure 1: Dr. Thomas David Anderson (1853 –1932)

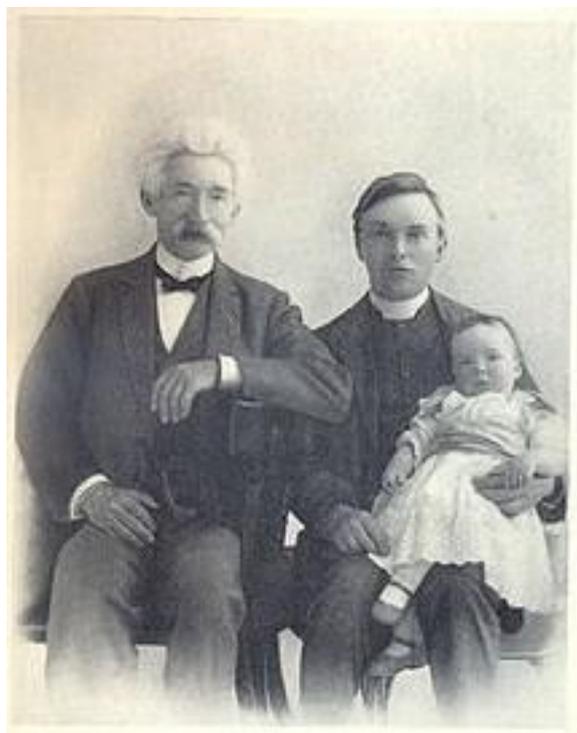

Figure 2: Dr. Hector Copland Macpherson FRSE, FRAS (1888–1956), right, his father Hector Carsewell Macpherson FRSE (1851–1924) and his infant son Hector Macpherson III (1923 – 1981)

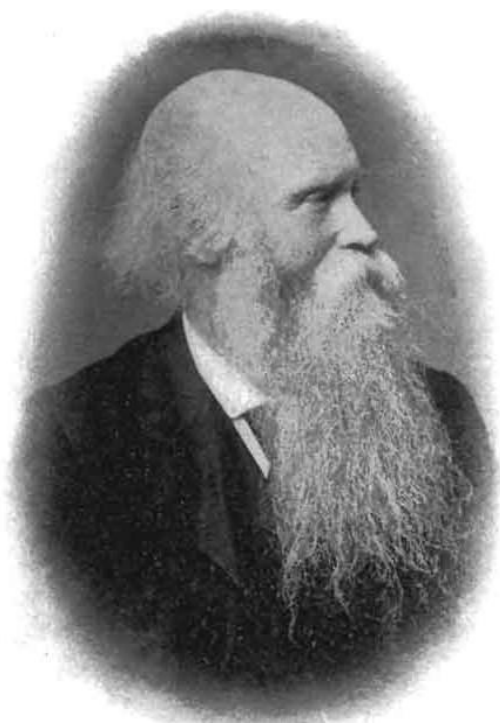

Figure 3: Prof. Ralph Copeland (1837-1905)

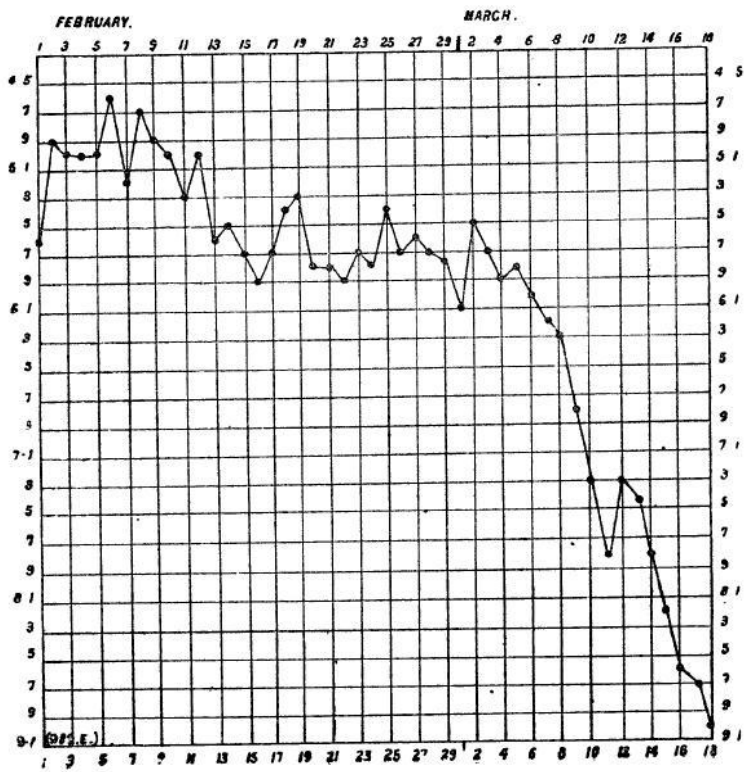

Figure 4: Light curve of Nova Aurigae 1891 during the 6 weeks after Anderson's detection.

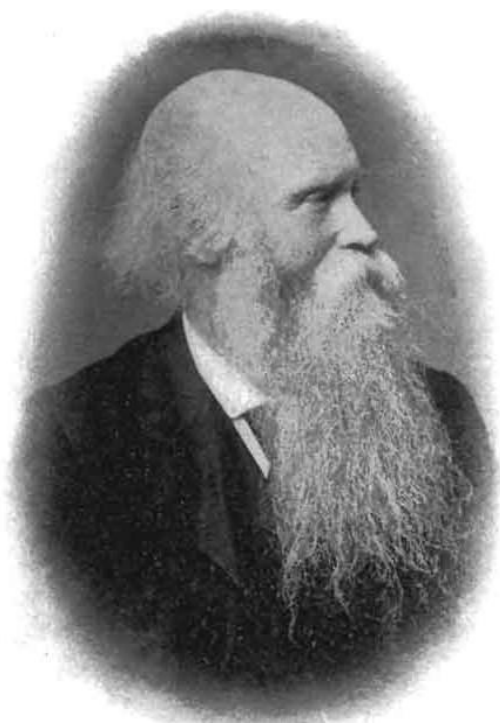

Figure 3: Prof. Ralph Copeland (1837-1905)

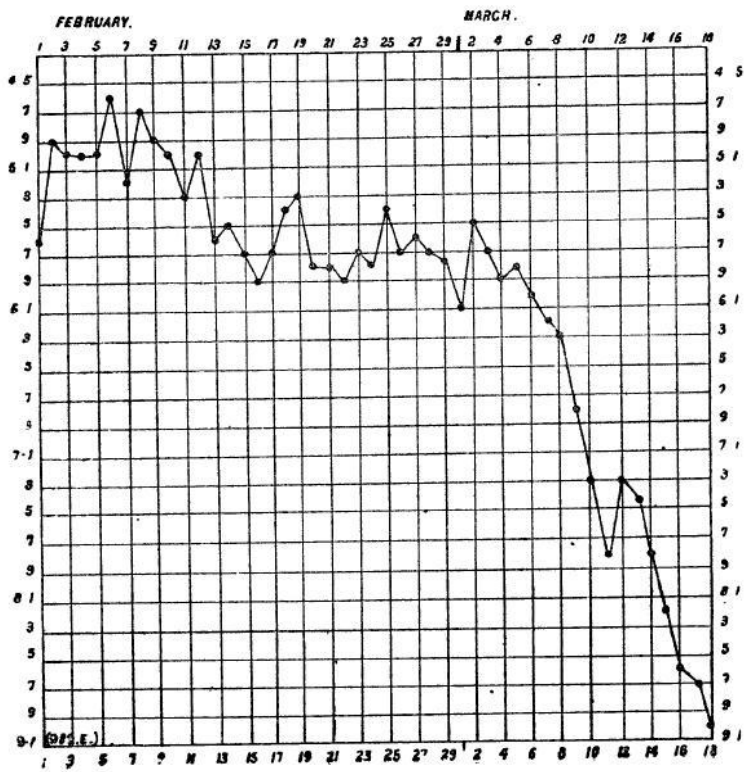

Figure 4: Light curve of Nova Aurigae 1891 during the 6 weeks after Anderson's detection.

Light curve drawn by Rev. T.H.E.C. Espin (70) showing daily means of data from himself, E.E. Markwick and George Knott, all of the BAA, and Paul Yendell, Edwin Sawyer, J. Plassmann, Reichwein and the Observatories at Oxford and Greenwich

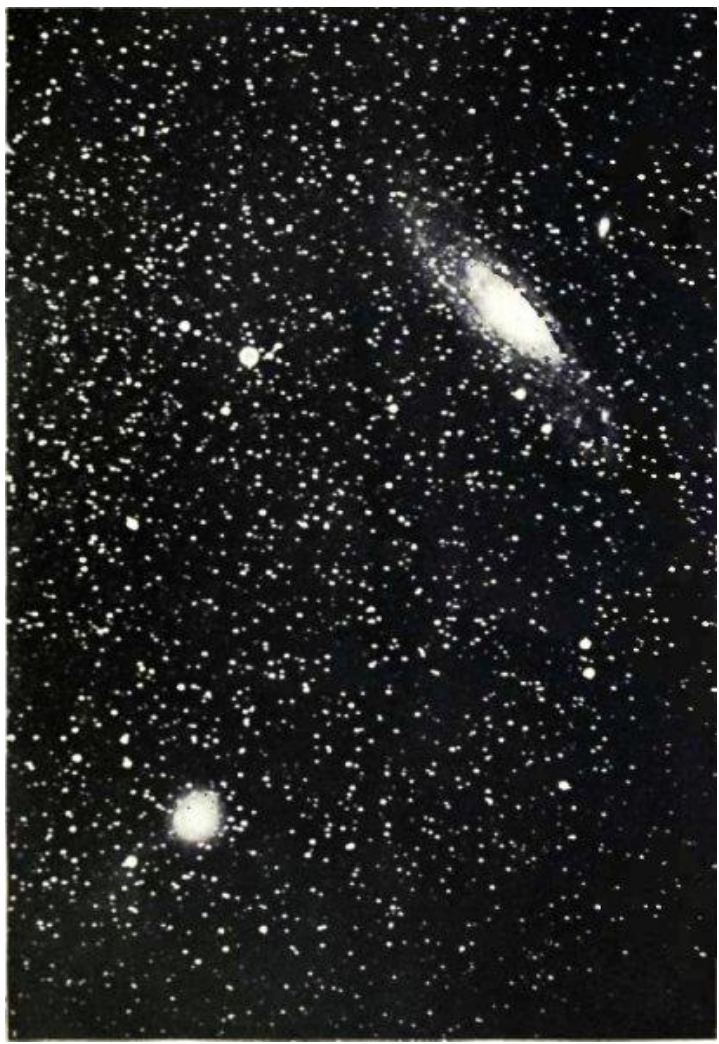

Figure 5: Comet 17P/Holmes and the Andromeda Galaxy, M31

Photograph by E.E. Barnard on 10 Nov 1892. From reference (71)

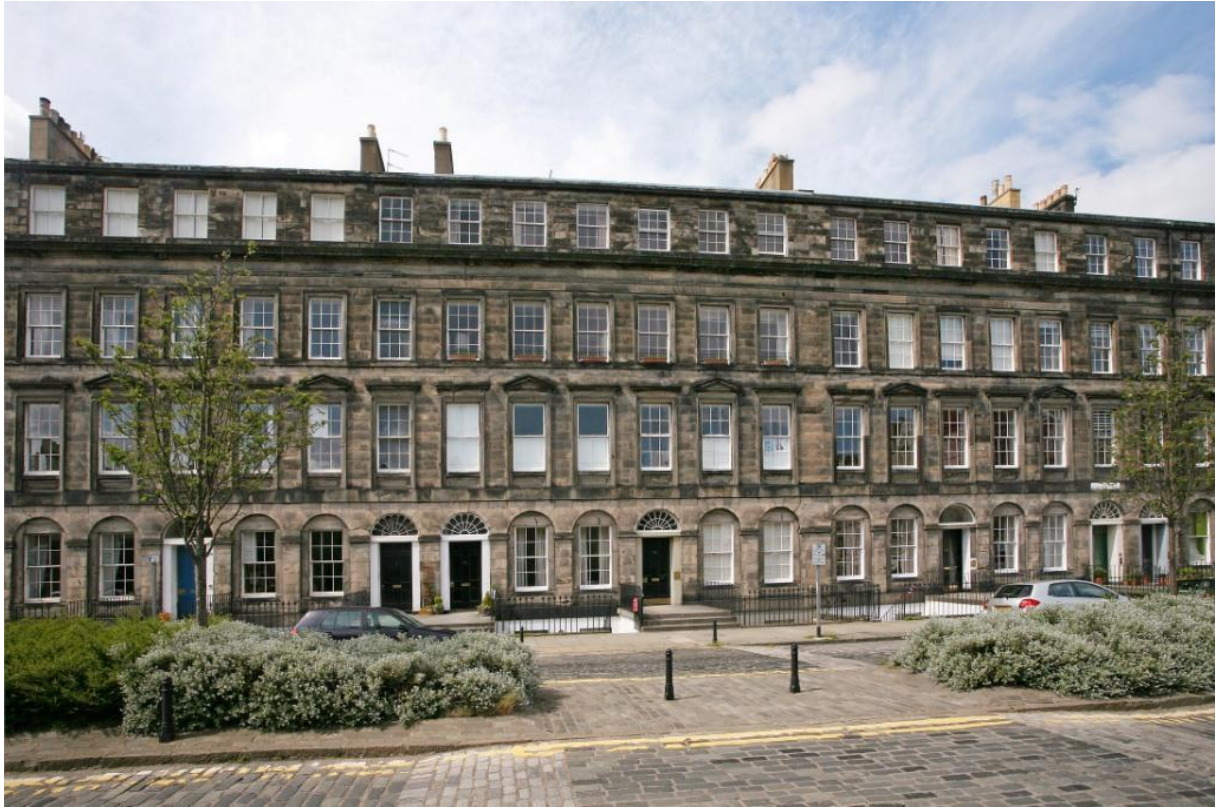

Figure 6: Anderson's home at 21 East Claremont Street Edinburgh

The entrance to Number 21 is the right hand of the pair of black doors to the left of centre.
Image courtesy of Simpson & Marwick, Edinburgh

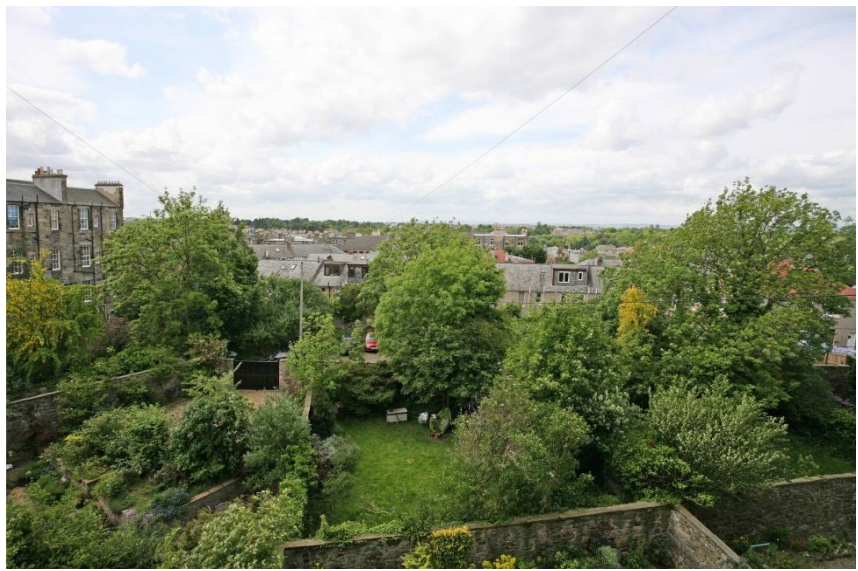

Figure 7: View from the rear of Anderson's home in Edinburgh

Image courtesy of Simpson & Marwick, Edinburgh

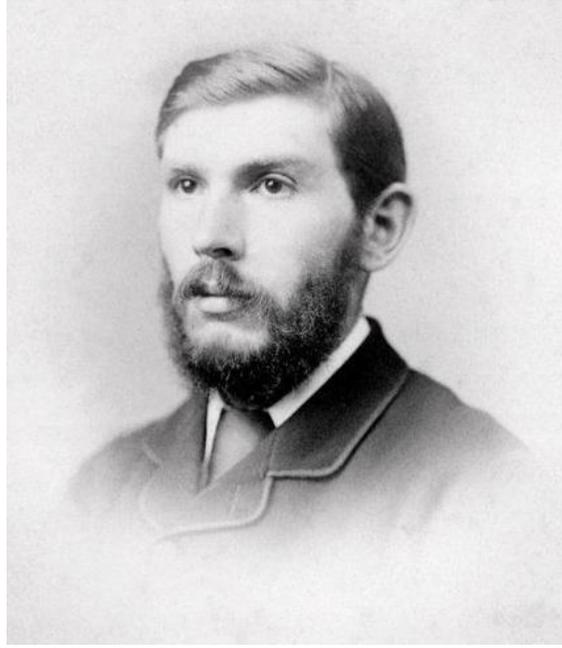

Figure 8: J.W.L. Glaisher (1848-1928)

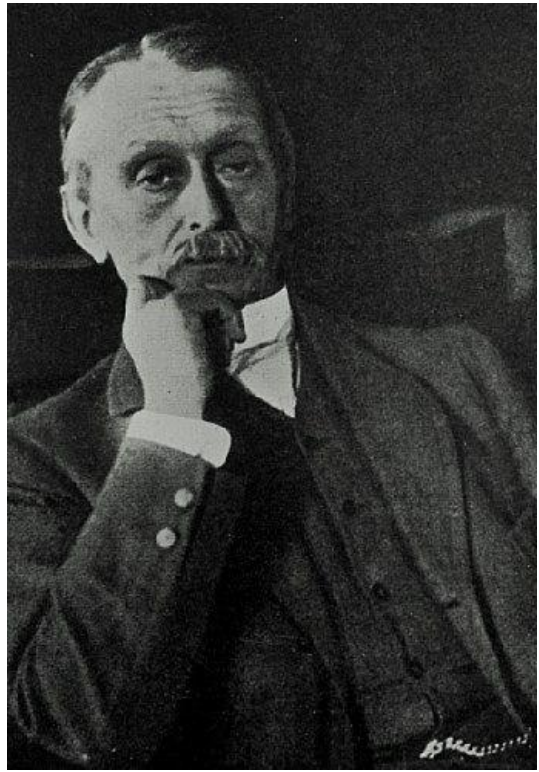

Figure 9: Prof. J.C. Kapteyn (1851 – 1922)

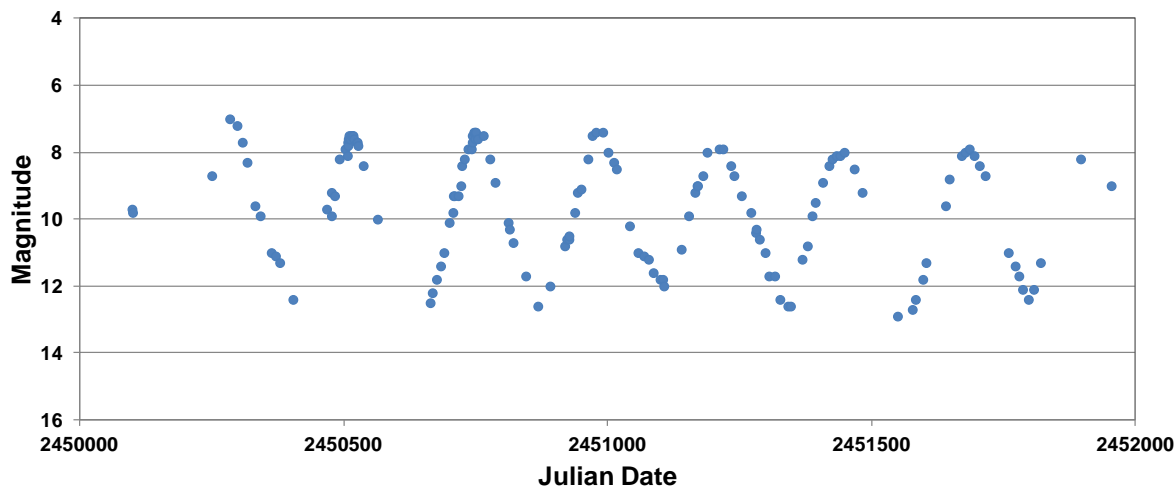

Figure 10: Visual light curve of V Cas between 9 October 1995 and 31 March 2001

Observers: A.R. Baransky, B.H. Granslo, P.J. Charleton, P. Veleshchuk, R.J. Bouma

(BAA VSS database)

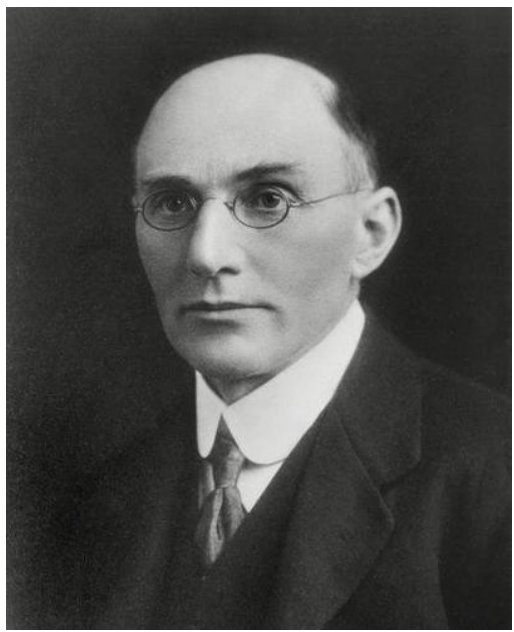

Figure 11: Prof. Ralph Allen Sampson (1866 – 1939)

RAS Presidential portrait

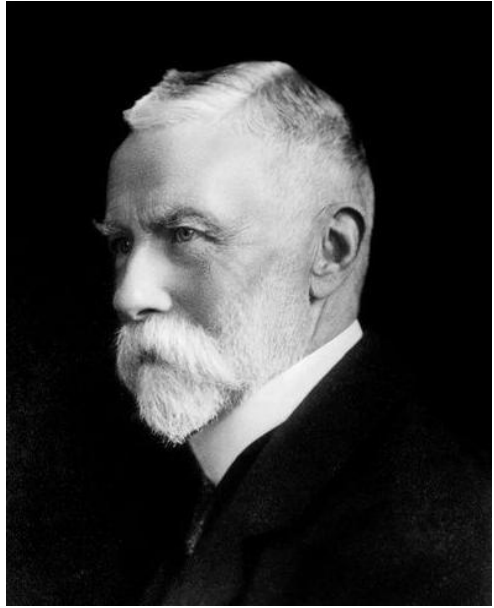

Figure 12: John Louis Emil Dreyer (1852-1926)

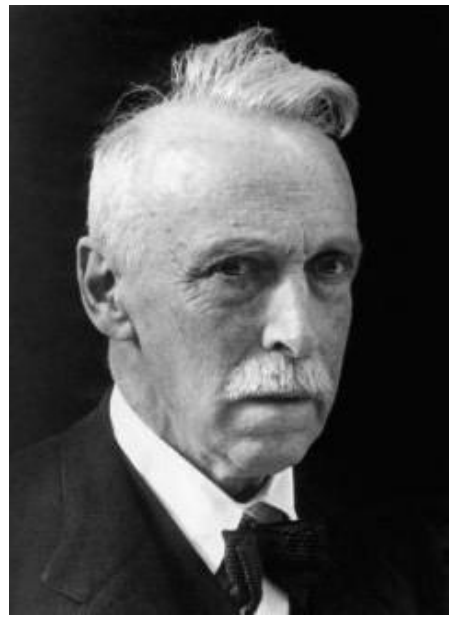

Figure 13: Ejnar Hertzsprung (1873-1967)